# Gradient-based adaptive wavelet de-noising method for photoacoustic imaging *in vivo*

Xinke Li,[1] Peng Ge,[1] Yuting Shen,[1] Feng Gao,[1] Fei Gao[1,2,3]

1 Hybrid Imaging System Laboratory, School of Information Science and Technology, ShanghaiTech University, Shanghai 201210, China

2 Shanghai Engineering Research Center of Energy Efficient and Custom AI IC, Shanghai, 201210, China

3 Shanghai Clinical Research and Trial Center, Shanghai 201210, China

**\* Correspondence**

Fei Gao, ShanghaiTech University, Huaxia Middle Road 393, 201210 Shanghai, China

Email: gaofei@shanghaitech.edu.cn

Photoacoustic imaging (PAI) has been applied to many biomedical applications over the past decades. However, the received PA signal usually suffers from poor signal-to-noise ratio (SNR). Conventional solution of employing higher-power laser, or doing long-time signal averaging, may raise the system cost, time consumption, and tissue damage. Another strategy is de-noising algorithm design. In this paper, we propose a new de-noising method, termed gradient-based adaptive wavelet de-noising, which sets the energy gradient mutation point of low-frequency wavelet components as the threshold. We conducted simulation, ex vivo and in vivo experiments to validate the performance of the algorithm. The quality of de-noised PA image/signal by our proposed algorithm has improved by 20%-40%, in comparison to the traditional signal denoising algorithms, which produces better contrast and clearer details. The proposed de-noising method provides potential to improve the SNR of PA signal under single-shot low-power laser illumination for biomedical applications *in vivo*.

## 1 | INTRODUCTION

Photoacoustic imaging (PAI), as a new hybrid imaging modality, combines the advantages of both optical and ultrasound imaging: the high contrast of optical imaging and good spatial resolution in deep tissue of ultrasound imaging[1]. The PA effect is induced when a light-absorbing object is exposed to a pulsed light source. When chromophores (such as melanin, hemoglobin, and water) absorb photons and undergo nonradiative relaxation, the temperature within the sample rises sharply, leading to thermoelastic expansion and the emission of ultrasonic wave. The broadband PA signal is detected by the ultrasound transducer, followed by signal processing and image reconstruction.

By utilizing the PA effect, PAI technology has been developed fast in recent decades, including two main categories: PA microscopy (PAM) and PA tomography (PAT), which targets different application scenarios requiring different penetration depth or spatial resolution[2]. By probing the endogenous chromophores mentioned above, PAI is able to image from organelles to organs in vivo[3, 4]. This new technology has been used in some preclinical and clinical applications, such as early detection for breast cancer, skin diseases, and so on[5-7].

However, there also exists many challenges in PAI technology for wide clinical use. One of the challenges is the low SNR of PA signal[8], especially in deep tissue imaging under limited laser energy. The underlying reason is that energy conversion efficiency from light to acoustic pressure, arising from the thermoelastic mechanism, is very low (ranging from $10^{-12}$ to $10^{-8}$)[9]. Additionally, acoustic attenuation and scattering in heterogeneous biological tissues further contribute to the problem. As a result, the received PA signal is often weak and strongly distorted, which results in blurring and artefacts in reconstructed PA image. If the biological tissue has multiple layers, the problem will be more complex. Due to the reflection, scattering and attenuation, the upper layer may generate PA signals with higher intensity and frequency spectrum, e. g. PA signal from the skin surface, which are usually unwanted signals. On the other hand, the lower layer may generate PA signals with lower intensity and frequency spectrum, which exhibits much worse SNR. To alleviate this problem, there usually exists two main approaches. The first approach is to increase the laser power, which is ultimately limited by the safety issue. The other approach is to do multiple signal acquisition and data averaging, which will severely slow the imaging speed[10].

In practice, signal preprocessing approaches, such as low-pass or band-pass filtering, is also widely used. It does filter out part of the noise, but it will also induce PA signal distortion and lose important details, leading to blurring of PA image. What's more, it performs even worse when dealing with multi-layer tissue structure, since different layer's PA signals usually shows different frequency characteristics. Wavelet threshold de-noising (WTD), as a potential signal processing method, has been applied in de-noising PA signals[11]. However, traditional WTD method often exhibit a dilemma: either excessive de-noising, causing distortion of the PA signals, or weak de-noising, resulting in inconspicuous noise reduction. Sometimes, it may go to the other end, inducing more impulse noise. In this paper, based on the frequency characteristics of the acquired PA signals, we propose a modified wavelet de-noising method, which can effectively reduce the noise and maintain the fidelity of PA signal. Both ex vivo and in vivo experiments were conducted to validate the feasibility of the proposed method.

# 2 | WAVELET THRESHOLD DENOISING

## 2.1 | Principle

Wavelet denoising method is a time-frequency analysis method based on the wavelet transform (WT) theory. It uses the characteristics of WT multi-resolution analysis, with less operational complexity.

Suppose the acquired PA signal $y(n)$ of length N is approximately in the following form:

$$y(n) = x(n) + \sigma e(n), \ 1 \le n \le N \quad (1)$$

where $x(n)$ is the clean PA signal, $e(n)$ is the noise, and $\sigma$ is the noise's standard deviation. The purpose of WTD is to recover $x(n)$ from $y(n)$ with as little distortion as possible[12].

In general, noise, mainly in the high frequency, has a larger number of wavelet coefficients with smaller magnitude, and the target signal, mainly in the low frequency, has a fewer number of wavelet coefficients with larger magnitude. According to this property, the basic idea of WTD is to set the wavelet coefficients below a certain threshold to zero, and preserve or shrink the wavelet coefficients above the threshold, corresponding to two coefficients reduction ways: hard thresholding and soft thresholding, respectively.

The main process of WTD is divided in the following three steps:

1. Select proper wavelet function and the number of decomposition layers. Then do wavelet decomposition to acquired PA signal.

2. Choose appropriate threshold. Do wavelet coefficient reduction, by either hard thresholding or soft thresholding.

3. Reconstruct PA signal based on the processed wavelet coefficients, by inverse wavelet transform[13].

There are also some trade-offs to be made during the de-noising process. The threshold can't be too large, or it will filter out the useful PA component. It can't be too small, either, or it will with relatively larger energy and adversely affect the de-noising performance. The same principle is also applicable for the number selection of decomposition layers. If the number of wavelet decomposition levels is too high, the useful PA signal may be compromised. Conversely, if the number of decomposition levels is too low, it becomes challenging to effectively distinguish between the signal and the noise.

## 2.2 | Traditional wavelet threshold selection method

The threshold selection is essential to the wavelet de-noising performance. There are many traditional threshold selection methods, such as sqtwolog, rigrsure, heursure, and minimaxi, which will be introduced below.

### 2.2.1 | Sqtwolog

Sqtwolog thresholding selection rule was proposed by Donoho and Johnstone in 1994[14]. The threshold $\lambda_i$ is calculated as

$$\lambda_i = \sigma_i \sqrt{2 \log N_i}, \ \sigma_i = \frac{MAD_i}{0.6745} = \frac{median(|\omega|)}{0.6745} \quad (2)$$

where $\sigma_i$ is the median absolute deviation (MAD), $N_i$ is the length of noisy signal and $\omega$ is the wavelet coefficients at $i_{th}$ scale.

### 2.2.2 | Rigrsure

Rigrsure is a soft threshold method evaluator of unbiased risk. It's calculated as follows:

$$\lambda_i = \sigma_i \sqrt{\omega_b} \quad (3)$$

$\omega_b$ represents the square of the $b^{th}$ wavelet coefficient (coefficient at minimal risk) selected from the vector $W = [\omega_1, \omega_2, ..., \omega_N]$. This vector comprises the squared values of the wavelet coefficients, arranged in ascending order. Here, $\sigma_i$ denotes the standard deviation of the noisy signal at $i_{th}$ scale.

### 2.2.3 | Minimaxi

Minimaxi is a fixed-value threshold selection method to find the optimal threshold for the root mean square error against the ideal procedure, which can be expressed as:

$$\lambda = \begin{cases} \sigma(0.3936 + 0.1829(\frac{lnN}{ln2})), & N > 32 \\ 0, & N \le 32 \end{cases} \quad (4)$$

Here, N is the signal length and $\sigma$ is the signal variance.

### 2.2.4 | Heursure

Heuisure is a threshold selecting method that combines both sqtwolog and rigrsure methods. If the SNR is small, the rigrsure method's estimation is poor. In that case, the fixed-form threshold from sqtwolog is used, and vice versa.

## 2.3 | Proposed gradient-based adaptive wavelet de-noising (gaWD) method

The above traditional wavelet threshold selection methods perform not so well in PA signal de-noising. Based on the characteristics of acquired PA signal, in this work, we propose a new gradient-based adaptive wavelet de-noising (gaWD) method. We first decompose the signal into 6 layers, using wavelet daubechies 8 (db8), shown in **FIGURE 1**(a). After careful evaluation, we find that D1, D2, and D3 layers are mainly noise layers. The coefficients with large magnitude among these three layers are not the target PA signal's main components. In the opponent, they are mainly from the large coupled signal at the transducer surface. The other coefficients in D1, D2 and D3 layers are usually much smaller than D4, D5, and D6 layers. D4, D5 and D6 layers are composed of both noise and signal components, where the magnitude of noise components is much lower than target PA signal. From D4 to D6, the wavelet coefficients go gradually closer to low frequency range, which means that the coefficients in D6 is more important than D4 and D5. What's more, the magnitude of wavelet coefficients represents signal energy, which means that from perspective of energy, we can get a threshold T from a more important layer D6, then apply it to D4 and D5, the two less important layers, ensuring filtering out all the coefficients with lower energy than target signal of D6 layer. Following this strategy, we sort D6 from smallest to largest, and name the new wavelet coefficients as D6_sort. After that, find the second largest gradient drop in D6_sort, which separates the target signal from the noise components. Then we set the corresponding coefficient value as threshold T. The whole threshold selection process is shown in **FIGURE 1**(b). After selecting the threshold, we further use hard thresholding method. For D1, D2, D3 layers, we set them all to zero. For D4, D5, D6 layers, we set all the coefficients below the threshold T to zero and preserve all the coefficients above the threshold T. The whole coefficients selection process is shown in **FIGURE 1**(c).

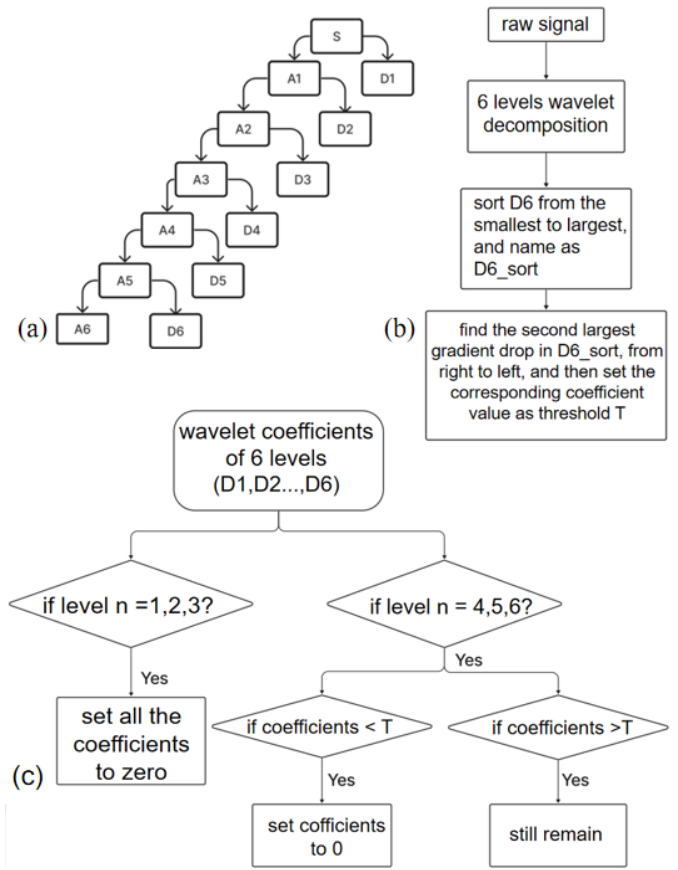

**FIGURE 1** The algorithm design of the proposed gaWD method. (a) The 6 levels of wavelet decomposition. (b) The process to find threshold T. (c) The whole process of de-noising method.

## 3 | SIGNAL DE-NOISING RESULTS

A typical acquired PA signal, from cancerous colorectal tissues, corrupted by noise and limited by resolution is shown in **FIGURE 2**. The final de-noising result using our proposed gaWD method, low pass filter, sqtwolog threshold, minimaxi threshold, hersure threshold, and rigrsure threshold method are shown in **FIGURE 3**, respectively.

From the signal de-noising results of 6 different methods shown in **FIGURE 3**, we can find that the classic wavelet de-noising methods, like sqtwolog, minimaxi, hersure, and rigrsure methods, all induce impulse noise distortion. They also fail to filter out the large coupled signal at the transducer surface. In comparison, sqtwolog threshold method performs better than the other 3 classic wavelet threshold methods, although it induces some signal distortion. Low pass filter and our proposed method both perform well in removing the coupled signal. However, the low pass filter preserves more noise in the waveform. On the contrast, our proposed

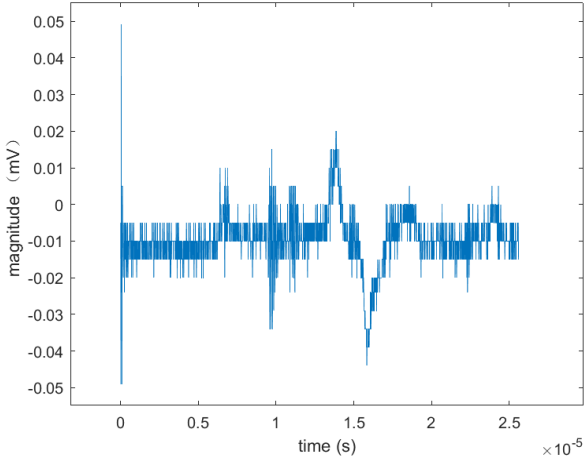

**FIGURE 2** The original signal corrupted by noise and limited by resolution

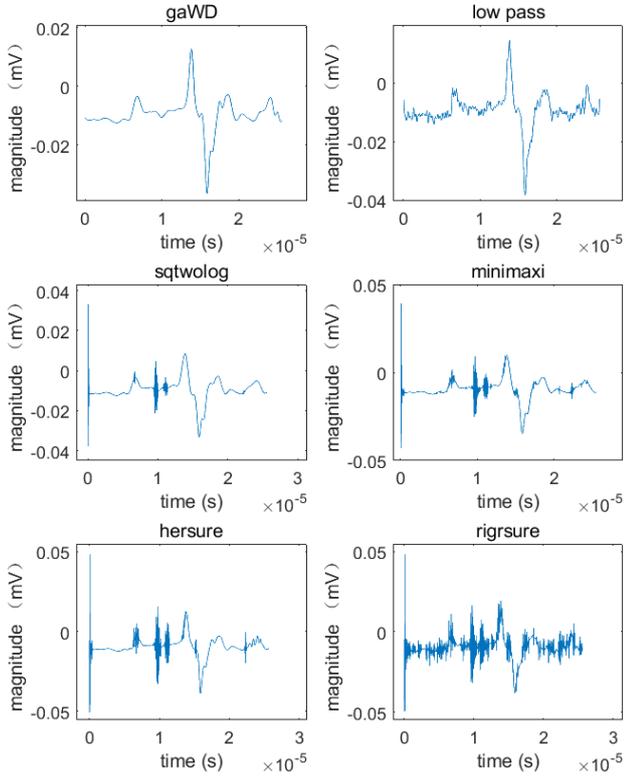

**FIGURE 3** The waveforms of de-noised signal after proposed gaWD method, low pass filtering, sqtwolog threshold, minimaxi threshold, hersure threshold, and rigrsure threshold method, respectively.

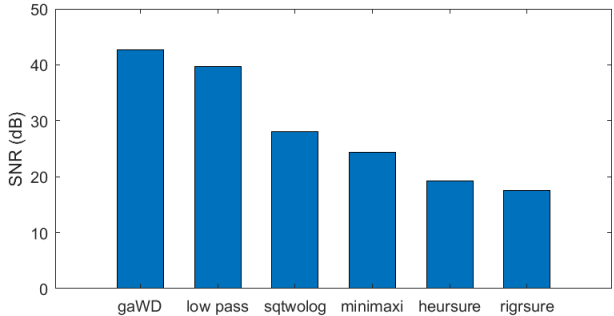

**FIGURE 4** The SNR of de-noised signal after proposed gaWD method, low pass filtering, sqtwolog threshold, minimaxi threshold, hersure threshold, and rigrsure threshold method, from left to right, respectively.

gaWD method achieves a much cleaner signal with higher SNR.

**FIGURE 4** shows the calculated SNR of the 6 different de-noising methods mentioned above. Our proposed gaWD method exhibits the highest SNR, about 3 dB above the low pass method, and much higher than the others.

## 4 | PA IMAGING SIMULATION RESULTS

In order to further verify our proposed algorithm, we use the k-wave toolbox in MATLAB to perform PA simulation study. Sixty-four acoustic sensor elements are set in a circle to receive PA signals generated from a segment of blood vessel. Since the sensors' distribution is sparse, the image will be blurred if we directly reconstruct the image. Here we first interpolate the recorded data on a continuous measurement surface before image reconstruction. The original distribution of both the vessel and sensors are shown in **FIGURE 5**(a). The reconstructed initial pressure distribution using interpolated data, without adding any noise, is shown in **FIGURE 5**(b). Then, we add 18dB noise to the raw PA data, leading to a noisy PA image in **FIGURE 5**(c). We perform 4-order Butterworth filtering, sqtwolog WTD, and our proposed method. The imaging results are shown in **FIGURE 5**(d), (e), (f), respectively. We can easily find that 4-order Butterworth filter blurs the vessel along with limited de-noising performance. The sqtwolog WTD de-noising method reduce much noise, but it also causes severe vessel signal distortion. In comparison, our proposed method achieves better de-noising performance without obvious signal distortion.

To be more quantitative, the PSNR and SSIM of these images are calculated according to the following equations (5)-(8) [15-17].

$$\mathbf{MSE} = \frac{1}{mn} \sum_{i=0}^{m-1} \sum_{j=0}^{n-1} ||f(i,j) - g(i,j)||^2 \tag{5}$$

$$\mathbf{PSNR} = 10\log_{10}\left(\frac{MAX_f^2}{MSE}\right) \tag{6}$$

$$\mathbf{SSIM(f,g)} = l(f,g)c(f,g)s(f,g) \tag{7}$$

$$\begin{cases} l(f,g) = \dfrac{2\mu_f\mu_g + C_1}{\mu_f^2 + \mu_g^2 + C_1} \\ c(f,g) = \dfrac{2\sigma_f + \sigma_g + C_2}{\sigma_f^2 + \sigma_g^2 + C_2} \\ s(f,g) = \dfrac{\sigma_{fg} + C_3}{\sigma_f\sigma_g + C_3} \end{cases} \tag{8}$$

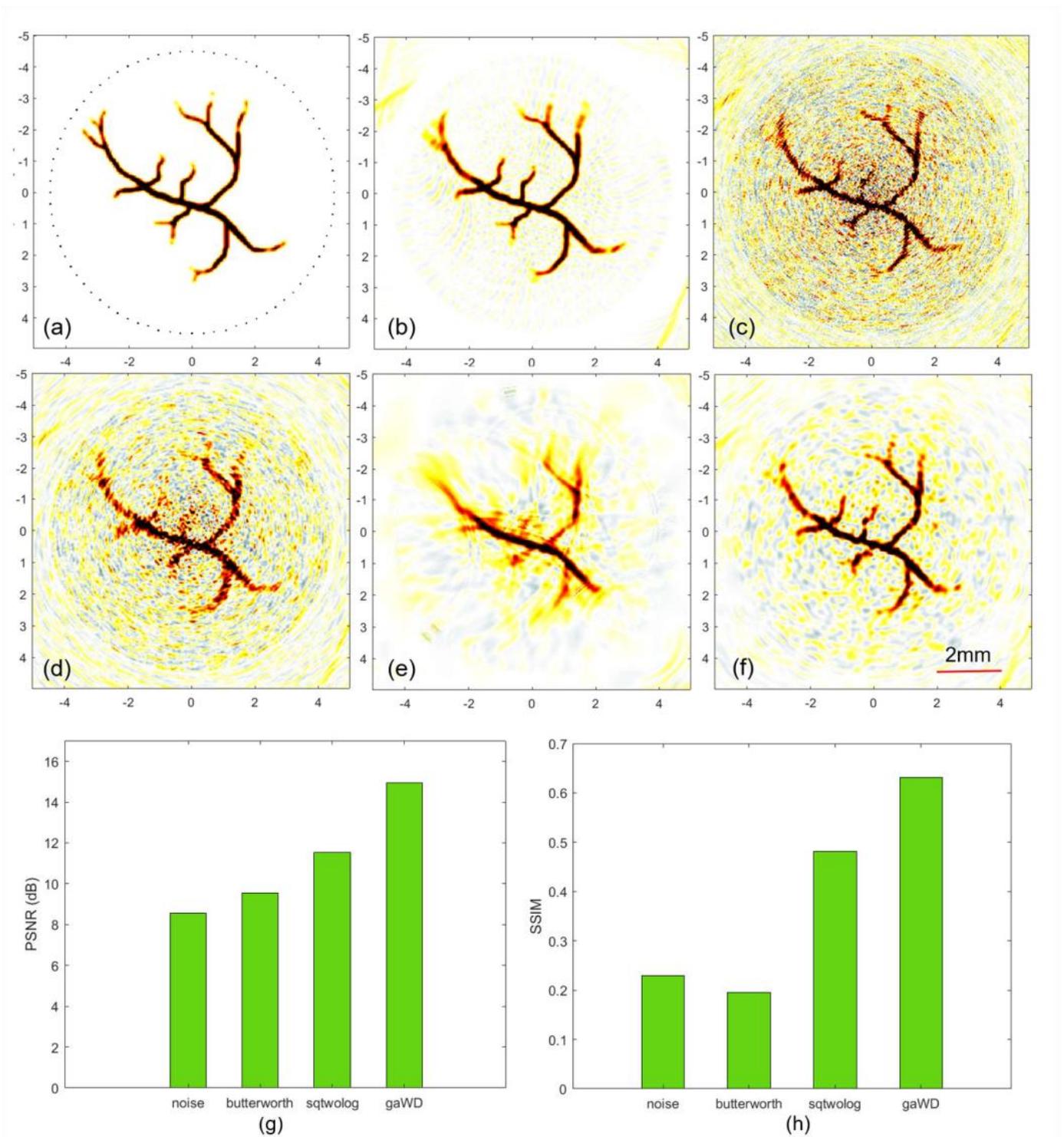

**FIGURE 5** (a) The initial pressure and sensor distribution in k-wave simulation. (b) The reconstructed PA image using interpolated data, without adding noise. (c) The reconstructed PA image using interpolated data, with 18 dB added noise. (d) The de-noised imaging results by Butterworth low pass filter. (e) The de-noised imaging results by sqtwolog threshold WTD method. (f) The de-noised imaging results by our proposed method (gaWD). (g) Quantitative comparison of PSNR of (c)-(f). (h) Quantitative comparison of SSIM of (c)-(f).

We set **FIGURE 5** (b) as our reference image. The calculated PSNR for comparison is shown in **FIGURE 5** (g). Similarly, the calculated SSIM of (c)-(f) is shown in **FIGURE 5** (h).

From **FIGURE 5**(g)-(h), we can find that our proposed gaWD method exhibits the highest PSNR and SSIM among all these methods.

# 5 | EXPERIMENTAL RESULTS

## 5.1 | Experimental system setup

We set up a photoacoustic microscopy (PAM) system to validate the effectiveness of our proposed de-noising algorithm. The schematic is shown in **FIGURE 6**. In this experimental setup, we focus the laser into a fiber by an objective lens (OL), and then use two cascaded lens to focus the output light of the fiber onto the sample. The received PA signal from ultrasound transducer (UT) is first fed into a low-noise amplifier (AMP) to be amplified with low noise induction, then transferred to data acquisition card for digitization, which is connected to a computer for real-time storage and display. The X/Y two-dimensional step motor is used for raster scanning. The data sampling rate is set to 80 MHz, and the transducer's central frequency is 10 MHz.

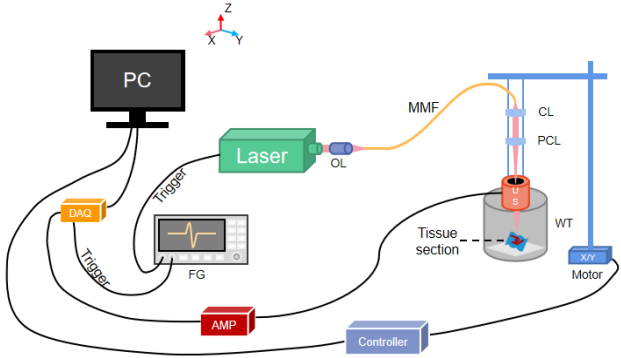

**FIGURE 6** The schematic of experimental system. FG: function generator; OL: objective lens; MMF: multimode fiber; CL: collimating lens; US: ultrasound; WT: water tank; AMP: amplifier; DAQ: data acquisition card; X/Y Motor: two-dimensional step motor.

## 5.2 | Ex-vivo experimental results

We conduct imaging of *ex-vivo* colorectal tissues first. We tried sqtwolog threshold method, 4-order Butterworth low pass filter, and our proposed gaWD method, respectively. We first draw the maximum amplitude projection (MAP) image after applying each de-noising method, shown in **FIGURE 7**. As shown in **FIGURE 7**(a), the MAP image reconstructed from raw data is severely corrupted by both random noise and stripe noise. It could be found that sqtwolog threshold method filters out some random noise, but is unable to reduce the stripe noise, shown in **FIGURE 7**(b). As for 4-order Butterworth low pass filtering result in **FIGURE 7**(c), it removes the stripe noise at the background, and enhances the image contrast. However, it blurs the image and loses some details, e.g., some texture of tissues is disappeared. Compared to these two methods, our proposed method achieves both de-noising and enhancing the image contrast, meanwhile preserving the details, presented in **FIGURE 7**(d). The corresponding SNR of all these images are calculated and shown in **FIGURE 7**(e). The de-noised image of our proposed method has the highest SNR, which validates the effectiveness of the proposed algorithm.

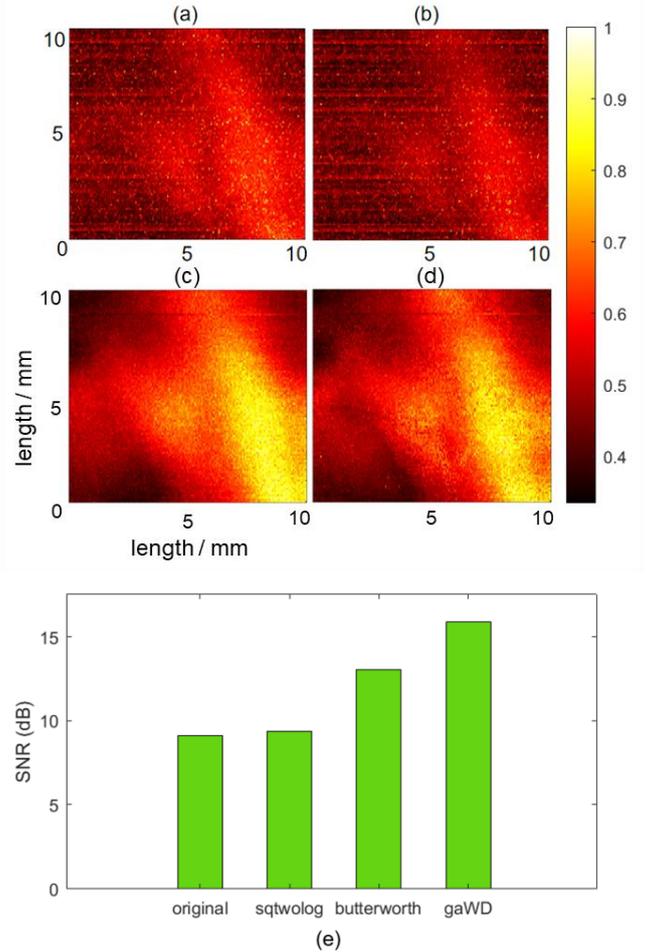

**FIGURE 7** The MAP image of ex-vivo colorectal tissues reconstructed from: (a) raw data, (b) de-noised data by sqtwolog threshold method, (c) de-noised data by 4-order Butterworth low pass filter, (d) de-noised data by our proposed de-noising method (gaWD). (e) the calculated SNR of (a)-(d).

We also plot the 3D volumetric PA image of the *ex-vivo* colorectal tissues, after de-noising processing by the above three methods, respectively. The results are displayed in **FIGURE 8**. Same with the de-noising results in **FIGURE 7**, the 3D volumetric PA image reconstructed by our proposed gaWD method shows cleaner in background, achieving higher image contrast and more detailed texture.

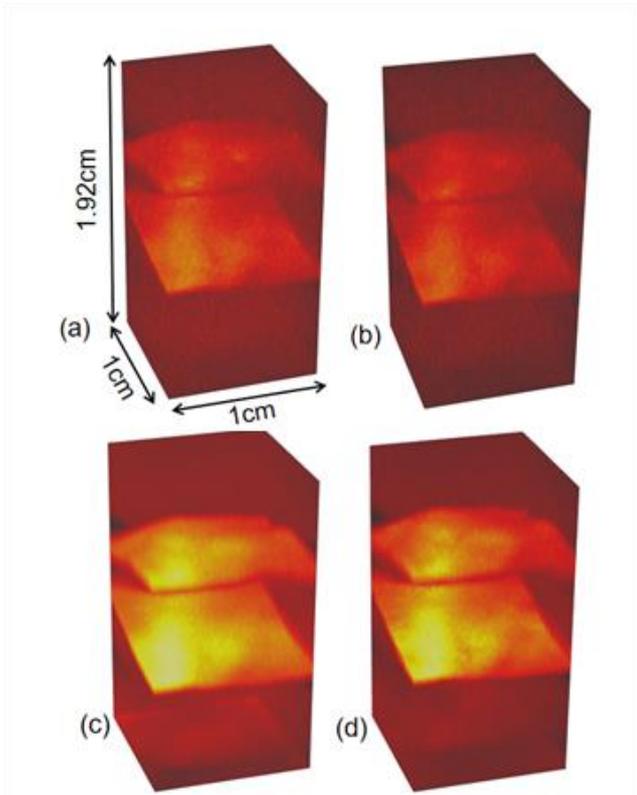

**FIGURE 8** The 3D volumetric image of ex-vivo colorectal tissues reconstructed from: (a) raw data, (b) de-noised data by sqtwolog threshold method, (c) de-noised data by 4-order Butterworth low pass filter, (d) de-noised data by our proposed gaWD method.

## 5.3 | In-vivo experimental results

Except for the *ex-vivo* colorectal tissues imaging, we also perform *in-vivo* blood vessels imaging of rat. To achieve deep penetration and good resolution for blood vessels simultaneously, we choose laser of 790 nm wavelength. Before experiment, we scrape the fur off the surface of the rat's skin at the target region, in order to reduce the scattering and attenuation of PA signal caused by the fur. The whole process of our experiment is strictly conducted under the standard of animal protection in animal research principles (Approving institution: ShanghaiTech University, IACUC protocol number: 20200323002). The photograph of the in vivo experimental setup is shown in **FIGURE 9**. To validate the algorithm's effectiveness in low SNR scenario, we add 18 dB noise to our raw data. We then process it by the above 3 methods, respectively. The 3D volumetric images are shown in **FIGURE 10**, respectively. First, we need to know there are two layers to be measured in this experiment. The skin surface of the rat generates large high-frequency PA signal, and the blood vessels about 1.3cm below the skin generates relatively low-frequency PA signal. **FIGURE 10**(a) shows the 3D image of raw data with added 18dB noise, both two layers are significantly blurred and submerged in noise. As shown in **FIGURE 10**(b), after sqtwolog denoising method, the background noise has been reduced a lot, but the image contrast is still low. As shown in **FIGURE 10**(c), we apply 4-order Butterworth filter to the data, which added 18dB noise. We can easily find that the PA signals of the two layers are obviously enhanced, but the important details and the whole skin surface layer appear blurred. Among the three de-noising methods, our proposed gaWD method performs best. It not only reduces the background noise, but also enhances the image contrast perfectly, without loss of useful information. Additionally, both the surface layer above and the blood vessels layer below are clearly recovered from the noise corrupted data. These results show that our proposed gaWD method exhibits good ability to preserve the signal details without inducing distortion, while filtering out the noise, for both the surface with high-frequency signals and inside tissues with low-frequency signals.

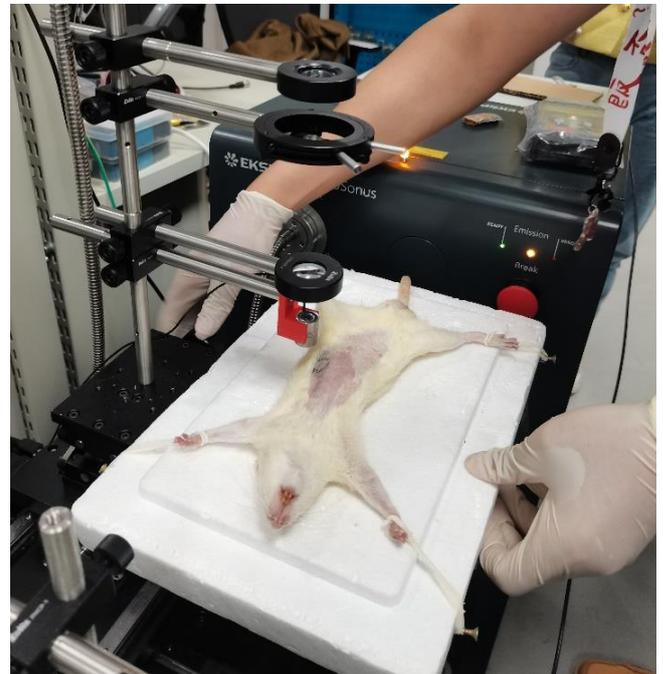

**FIGURE 9** The photograph of the rat placement for in vivo experiment.

## 6 | CONCLUSION

In conclusion, our proposed gradient-based adaptive wavelet denoising algorithm (gaWD) is validated to have good performance in noise reduction, details preservation, and image enhancement in PA imaging for both simulation, *ex-vivo* and *in-vivo* results. It has demonstrated superior performance compared to the other existing methods, like Butterworth filter, sqtwolog threshold denoising, and so on, with 20%-40% improvement. What's more, our proposed method is significantly effective in processing the two-layers structure, which exhibits different PA signal's frequency band. In such situation, our proposed method performs better than the existing de-noising methods. By employing the proposed fast and high-performance de-noising algorithm, it enables the using of a lower power laser to achieve the same SNR, which will decrease the cost of a PA imaging system.

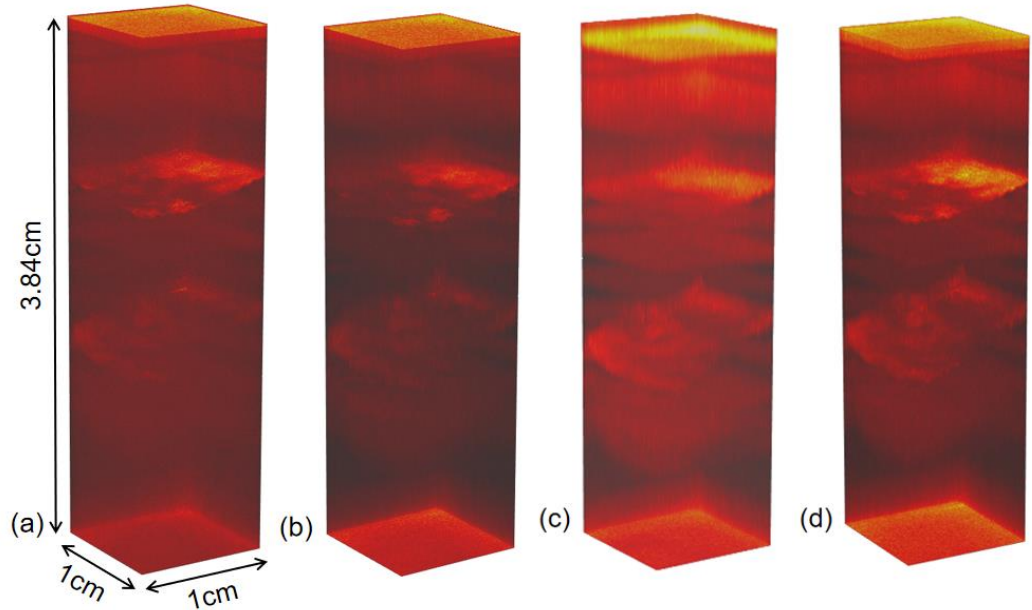

**FIGURE 10** The 3D volumetric image of *in-vivo* blood vessels image. (a) the raw data added 18dB noise. (b) the denoised data by sqtwolog method. (c) the denoised data by a 4-order butterworth lowpass filter. (d) the denoised data by our new proposed method (gaWD).